\documentclass[conference]{IEEEtran}
\IEEEoverridecommandlockouts
\usepackage{cite}
\usepackage{amsmath,amssymb,amsfonts}
\usepackage{graphicx}
\usepackage{textcomp}
\usepackage{xcolor}
\usepackage{amsmath, amssymb}
\usepackage{algorithm}
\usepackage{algpseudocode}
\usepackage{physics}
\usepackage{float} 
\usepackage{subcaption}
\usepackage{booktabs}
\usepackage{multirow}
\usepackage{wrapfig} 
\usepackage{array}    % for column width adjustments
\usepackage{booktabs} % cleaner lines
\usepackage{caption}  % for tight caption spacing
\usepackage{adjustbox}
\usepackage{graphicx}
\usepackage{soul}
\usepackage{makecell}
\usepackage{booktabs} % Required for \addlinespace and rules

\def\BibTeX{{\rm B\kern-.05em{\sc i\kern-.025em b}\kern-.08em
    T\kern-.1667em\lower.7ex\hbox{E}\kern-.125emX}}
\begin{document}
\bstctlcite{IEEEexample:BSTcontrol}

\title{Resilient Control Loops in Autonomous Vehicles Under Adversarial Jamming via Spectral Perception and Network-Layer Failover\thanks{This paper has been accepted for presentation and publication at the 2026 IEEE World Forum on Public Safety Technology (WF-PST).}}

% possibly "Resilient Autonomous Vehicle Control Loops Under Adversarial Interference via Spectral Perception and Network Failover"
%\title{Experimental Intentional Electromagnetic Interference (IEMI) Detection and Active Mitigation for Autonomous Vehicles}

% \author{
%     Luis Barajas,~\IEEEmembership{Student Member,~IEEE},
%     Colin Jeardoe,~\IEEEmembership{Student Member,~IEEE}\\
%     Jaewon Kim,~\IEEEmembership{Member,~IEEE}, 
%     Eman Hammad,~\IEEEmembership{Senior Member,~IEEE}
% }

% \author{
%   \IEEEauthorblockN{
%     Luis Barajas\IEEEauthorrefmark{1}\IEEEauthorrefmark{3}, 
%     Colin Jeardoe\IEEEauthorrefmark{1}\IEEEauthorrefmark{3}, 
%     Jaewon Kim\IEEEauthorrefmark{2}, 
%     and Eman Hammad\IEEEauthorrefmark{1}\IEEEauthorrefmark{3}
%   }
%   \IEEEauthorblockA{
%     Texas A\&M University, College Station, TX, USA \\
%     \IEEEauthorrefmark{1}iSTAR Laboratory,  
%     \IEEEauthorrefmark{2}Global Cyber Research Institute (GCRI) \\
%     \IEEEauthorrefmark{3} Department of Engineering Technology and Industrial Distribution \\
%     Emails: \{luisfba, cjeardoe, j1k, eman.hammad\}@tamu.edu
%   }
% }

\author{
  \IEEEauthorblockN{
    Luis Barajas\textsuperscript{1,3}, 
    Colin Jeardoe\textsuperscript{1,3}, 
    Jaewon Kim\textsuperscript{2}, 
    and Eman Hammad\textsuperscript{1,3}
  }
  \IEEEauthorblockA{
    Texas A\&M University, College Station, TX, USA \\
    \textsuperscript{1}iSTAR Laboratory,  
    \textsuperscript{2}Global Cyber Research Institute (GCRI) \\
    \textsuperscript{3}Department of Engineering Technology and Industrial Distribution \\
    Emails: \{luisfba, cjeardoe, j1k, eman.hammad\}@tamu.edu
  }
}

\maketitle

%%%%%%%%%%%%%%%%%%%%%%%%%%%%%%%%%%%%%%%%%%%%%%%%%%%%%%%%%%%%%%%%%%%%%%%%%%%%%%%%%%%%%%%%%%%%%%%%%%%%%%%%%%%%%%%%%%%%%%%%%%%

\begin{abstract}

The operational integrity of autonomous mobile robots relies on the continuous availability of wireless control loops, making them highly attractive targets for adversarial intentional electromagnetic interference. This paper introduces a resilient, cross-layer framework that combines physical-layer spectral perception with network-layer routing optimization to protect middleware stability, such as ROS~2, during intentional electromagnetic interference. Utilizing a software-defined radio front-end, the system extracts dynamic spectral descriptors, including spectral entropy and channel occupancy, to inform a Random Forest classifier that establishes adaptive environmental baselines. To ensure uninterrupted data flow, the architecture maintains dual pre-authenticated physical interfaces in a hot-standby configuration, enabling instantaneous failover through automated network routing table updates. Empirical validation on a physical ROS~2 mobile robot testbed demonstrates that this adaptive hardware-assisted architecture optimizes communication recovery to an average of 141~ms. This sub-second restoration translates directly into a 78.9\% reduction in pooled root-mean-square path tracking error compared to software re-association, successfully securing system-level mission integrity.

\end{abstract}


%%%%%%%%%%%%%%%%%%%%%%%%%%%%%%%%%%%%%%%%%%%%%%%%%%%%%%%%%%%%%%%%%%%%%%%%%%%%%%%%%%%%%%%%%%%%%%%%%%%%%%%%%%%%%%%%%%%%%%%%%%%%%

\section{Introduction}
\label{sec:intro}

The rapid proliferation of autonomous intelligent vehicles (UxVs) across commercial and industrial sectors has created a critical dependency on wireless communication for telemetry, navigation, and coordination~\cite{farraj2024physical}. However, this reliance on RF channels renders these systems highly susceptible to Intentional Electromagnetic Interference (IEMI) and jamming attacks \cite{10876697}, as evidenced by recent global events involving UAV losses to electronic warfare and widespread GPS disruptions \cite{Ferreira2020GPSJamming}. Despite progress in wireless robotics, maintaining the low-latency communication required for stable feedback loops remains a significant challenge; RF interference induces latency and packet loss that directly degrades real-time control timing. While current research often focuses on theoretical link-layer metrics \cite{ma2024against, wang2024reinforcement, alcorn2025situational, alcorn2026darrms}, there remains a critical gap in empirical evaluations of how physical-layer interference propagates through middleware, such as ROS 2, to disrupt the stability of physical robotic platforms.
\begin{figure}[!htbp]
  \centering
  \includegraphics[width=0.8\columnwidth]{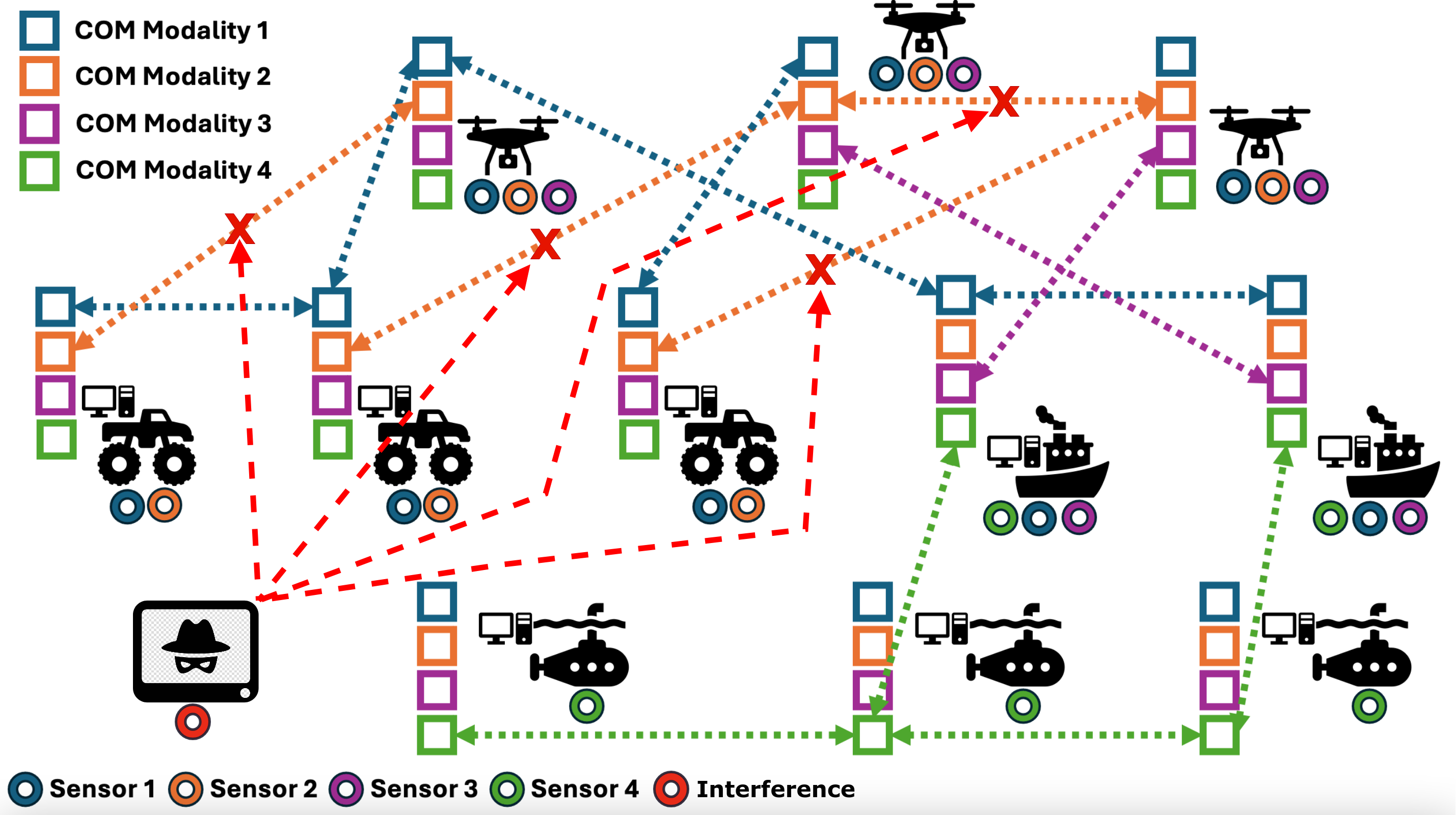}
  \caption{Example interference scenario illustrating how environmental noise and intentional interference can degrade UxV communication.}
  \label{fig:overview}
  \vspace{-15pt}
\end{figure}

Environmental variability further complicates resilience. Factors such as background noise and multipath reflections introduce significant uncertainty, meaning fixed detection thresholds often result in false alarms or missed detections \cite{chiper2022drone}. To maintain awareness, an autonomous vehicle must distinguish between ordinary environmental noise and intentional attacks by continuously monitoring spectrum features like the ambient noise floor. Furthermore, modern jamming has evolved beyond broadband noise; attackers now utilize adaptive techniques to target specific protocols. This escalation necessitates defensive systems that move beyond predefined patterns toward active, context-aware reasoning and recovery.

This paper addresses these challenges by framing communication resilience as a cross-layer agentic AI problem and providing a framework that integrates real-time SDR sensing with multi-modal communication switching for enhanced situational awareness. We develop an adaptive machine learning-based detection method that monitors spectral features to distinguish between environmental fluctuations and intentional interference, implemented within a proof-of-concept testbed using HackRF and USRP B210 radios. By evaluating this system on ROS 2-based UxVs, we provide empirical results quantifying the impact of interference on closed-loop control; our findings demonstrate that hardware-assisted autonomous band switching achieves a recovery latency of 141 ms and a 78.9\% reduction in pooled RMS path error compared to standard software-based methods.

% %%%%%%%%%%%%%%%%%%%%%%%%%%%%%%%%%%%%%%%%%%%%%%%%%%%%%%%%%%%%%%%%%%%%%%
% 
\section{Background and Related Works}
\label{sec:related_works}

%\subsubsection{RF Interference and Intentional Electromagnetic Interference (IEMI) on Autonomous Platforms}
Reliable wireless communication is critical for autonomous unmanned systems; however, it remains one of the most fragile components in real-world deployments. Recent incidents highlighted the urgency of this problem; where the ongoing Russia–Ukraine conflict have documented unprecedented UAV attrition rates attributable to electronic warfare (EW), with estimates of losses on the order of 10,000 drones per month, predominantly due to jamming and GPS spoofing \cite{Ukraine_drone, jhanjhi2025rise}. \cite{parlin2018jamming} demonstrated that protocol-aware jammers implemented on low-cost SDR platforms can disrupt mission-critical UAV remote-control links at significantly lower jam-to-signal ratios than naive broadband jammers, establishing the feasibility of asymmetric IEMI threats against unmanned systems. 

\subsubsection{SDR-Based Experiments and Counter-UAV Systems}
The proliferation of affordable, programmable software-defined radios (SDR), including the HackRF One, BladeRF, and the NI USRP family, has lowered the barrier for adversaries and enabled repeatable interference experimentation. \cite{chiper2022drone} presented a survey of SDR-based drone detection and defense systems, establishing taxonomies for both passive sensing and active counter-UAV architectures. \cite{perotoni2025electronic} characterized SDR-generated GPS interference against geo-location subsystems and quantified the receiver sensitivity thresholds at which navigation lock is lost, while \cite{alamleh2024system} proposed an integrated RF-SDR framework that fuses detection with selective jamming for fine-grained spectrum enforcement in restricted zones.

\subsubsection{ML-Based Jamming Detection and Classification}

Recent works have applied machine learning (ML) to jamming detection, motivated by the inadequacy of static-threshold detectors under variable RF environments. \cite{li2022jamming} introduced a feature-based and spectrogram-tailored ML pipeline that classifies four jammer types (barrage, single-tone, successive-pulse, and protocol-aware) against OFDM-based UAVs, demonstrating that spectral entropy and signal energy descriptors are discriminative features even under low signal-to-jamming ratios. \cite{price2022real} leveraged OFDM, energy, and SNR features for receiver-side inference. Deep learning methods were applied to cognitive radio spectrum sensing \cite{roopa2024deep} and to large-scale wideband signal identification, illustrating how convolutional architectures can generalize across modulation and interference. More recently, reinforcement learning (RL) emerged as a dominant paradigm; where \cite{ma2024against} surveyed RL-based anti-jamming strategies, and \cite{wang2024reinforcement} proposed an RL-driven jamming detector with explicit reliability guarantees. %While these methods primarily target the physical or link layer, few expose their inferences to the middleware or behavioral layer of the autonomous platform, which is a gap that our work addresses by routing detector outputs into the robot's higher-level decision logic.

\subsubsection{ROS 2 and DDS Middleware Resilience}
The Robot Operating System 2 (ROS 2) has become the middleware for modern autonomous systems, with its publish–subscribe communication built atop the Data Distribution Service (DDS) standard. \cite{maruyama2016exploring} first characterized the latency and throughput properties of ROS 2 under nominal conditions, and \cite{kronauer2021latency} subsequently decomposed end-to-end latency into ROS 2 core and DDS middleware contributions. \cite{kim2018security} analyzed the security-performance trade-offs introduced by the DDS Security specification and identified inconsistencies between the specification and reference implementations. In \cite{lee2025probabilistic} authors propose probabilistic latency analyses to quantify how loss ratios propagate into jitter and message-delivery rate over lossy wireless links, while \cite{lee2025optimizing} specifically examines ROS 2 communication optimization for wireless robotic deployments. \cite{paul2024performance} further evaluated multiple DDS implementations under cooperative-driving workloads, identifying both QoS settings and the Linux network stack as latency bottlenecks. However, these works uniformly treat the wireless channel as an abstract loss and latency source; they do not investigate how physical-layer IEMI propagates through DDS heartbeats, retransmission timers, and QoS policies to disturb closed-loop control. Our work bridges this gap by coupling SDR-generated IEMI with ROS 2-mediated waypoint tracking on a physical testbed.

\subsubsection{Adaptive Anti-Jamming and Multi-Modal Switching}

\color{black}
Prior anti-jamming work falls into three families. Frequency-domain evasion
relocates the link away from the jammed channel through proactive or reactive
channel hopping and rate adaptation~\cite{navda2007,djuraev2017channel,hanawal2015joint};
learned detectors classify interference from spectral features, as in
feature-based classifiers for OFDM-UAV links~[12],~[13]; and
reinforcement-learning and game-theoretic methods learn evasion policies
against adaptive jammers~\cite{yang2025agent,yin2024uav,pourranjbar2021reinforcement,cheng2025deep,qin2025multi,lin2024reinforcement}.
The present work differs on two axes. First, it recovers by switching between
two pre-authenticated interfaces at the network layer, avoiding the
single-interface re-association latency these schemes incur. Second, whereas
prior work is largely simulation-bound and evaluated by link-layer metrics, we
quantify the defense by its effect on a physical robot's closed-loop
path-tracking error. Like all evasion-based defenses, band switching can be
defeated by reactive or persistent jammers that follow the link across
bands~\cite{wilhelm2011reactive,lee2014persistent}, an adversary class the
continuous-wave evaluation here does not exercise and that motivates the
broadened threat model in 
Section~\ref{sec:futurework}.
 
\color{black}

\section{Methodology and Approach}
\begin{figure}[!htbp]
  \centering
  \includegraphics[width=0.98\columnwidth]{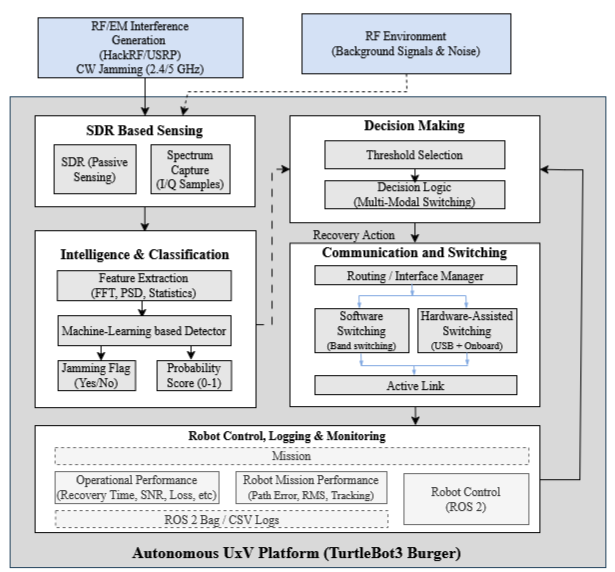}
  \caption{IEMI detection and recovery system overview.}
  \label{fig:DiagramTest.png}
\end{figure}
The proposed approach frames resilience as a cross-layer synchronization problem. The methodology is divided into two phases: (i) an environment-aware detection phase that distinguishes between natural noise and intentional interference, and (ii) a network-layer mitigation phase that leverages physical layer redundancy to maintain control-loop stability.

\noindent \subsubsection{Context-Aware Interference Perception} A fundamental challenge in communication resilience is the highly dynamic nature of the RF environment. Hence, instead of fixed detection thresholds we adopt a feature-based classification strategy that treats the local spectrum as a dynamic input to the system's decision-making process. By extracting spectral features such as spectral entropy, channel occupancy, and power spectral density (PSD) variance, the system continuously characterizes the nominal state of the environment, enabling adaptation to shifting noise floors caused by mobility or obstacles. When observed features deviate significantly from this learned nominal baseline, the framework identifies an IEMI event. This classification if fed to the application layer, providing the robotic agent with the cross-layer situational awareness required to trigger recovery.

\noindent \subsubsection{Multi-Modal Path Redundancy} The mitigation strategy is built upon the principle of path redundancy. Traditional software-based recovery is inherently slow because it relies on a single radio interface to cycle through deauthentication and re-association phases. Our methodology proposes an architecture where multiple authenticated communication pathways are maintained simultaneously across diverse physical layers. While this work validates the concept using dual-band Wi-Fi (2.4 GHz and 5 GHz), the approach is designed to be agnostic to the specific protocol and can be extended to redundant links across cellular (5G), satellite, or proprietary sub-GHz radios. By maintaining hot-standby connections, the system eliminates the temporal bottleneck of the link-layer handshake.

\noindent \subsubsection{Network-Layer Mitigation and Control Loop Stability} For seamless mission continuity, the transition between communication paths is handled at the network layer rather than the link layer. By modifying the system’s routing table and default gateways in response to a classified threat, the framework redirects outbound telemetry and control traffic instantaneously. This architectural choice is critical for systems utilizing middleware like ROS 2, where the stability of the closed-loop control is highly sensitive to jitter and packet loss. By reducing the recovery interval from the multi-second range (standard software re-association) to the millisecond range (routing-table modification), the methodology ensures that the physical-layer disruption does not propagate into a catastrophic failure of the robotic motion task.

\noindent \subsubsection{Validation Case Study: Dual-Band Wi-Fi Testbed} The proposed approach is validated through a cyber-physical testbed. In this experiment, a ROS 2-based autonomous vehicle utilizes two independent Wi-Fi interfaces associated with different frequency bands. This setup serves as a controlled environment to quantify how the proposed hardware-assisted switching reduces recovery latency (achieving 141 ms) and minimizes path-following errors (reducing RMS error by 78.9\%) compared to standard software-based methods.

%%%%%%%%%%%%%%%%%%%%%%%%%%%%%%%%%%%%%%%%%%%%%%%%%%%%%%%%%%%%%%%%%%%%%%%%%%%%%%%%%%%%%%%%%%%%%%%%%%%%%%%%%%%%%%%%%%%%%%%%%%%%%

%\clearpage

\section{Experimental Setup}
\label{sec:experimental_setup}

This section describes the hardware platform, software stack, wireless environment, and three switching architectures used to evaluate the proposed IEMI detection and recovery framework. The setup is organized to ensure reproducibility and to support the closed-loop evaluation methodology.

\subsection{Robot Platform and Software Stack}

The experimental platform was based on a TurtleBot3 Burger configured with its standard hardware architecture. The robot employed a Raspberry Pi 4 (8 GB RAM) running Ubuntu Server 24.04 as the onboard compute unit, paired with an OpenCR controller for low-level motor control. ROS 2 Humble Hawksbill served as the middleware layer for command, telemetry, and state estimation, with Cyclone DDS as the underlying Data Distribution Service (DDS) implementation. 

\begin{figure}[!htbp]
     \centering
     \begin{subfigure}[b]{0.46\columnwidth}
         \centering
         \includegraphics[width=\textwidth]{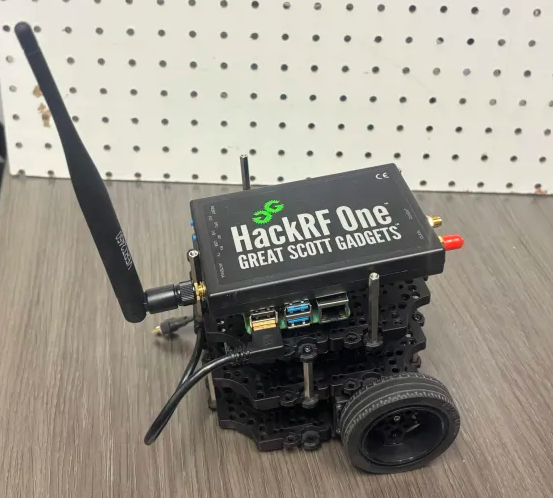}
         % \caption{Mobile robot platform, Turtlebot 3}
         \label{fig:siderobot}
     \end{subfigure}
     \hfill
     \begin{subfigure}[b]{0.48\columnwidth}
         \centering
         \includegraphics[width=\textwidth]{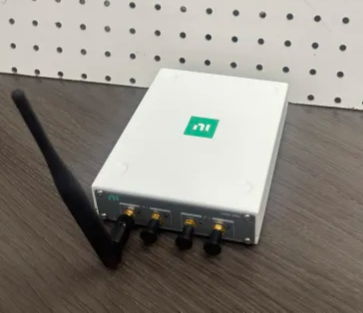} % Replace with your SDR filename
         % \caption{Software Defined Radio, USRB}
         \label{fig:USRP}
\end{subfigure}
\caption{Experimental setup components including (a) the mobile robot platform, and (b) the Software Defined Radios.}
\vspace{-10pt}
\end{figure}

\subsection{Software-Defined Radio (SDR) Platforms}

Two software-defined radio (SDR) platforms were used in the testbed: the HackRF One and the NI USRP B210. The HackRF One served as the robot-side passive sensing front end, while the USRP B210 served as the controlled interference transmitter. The HackRF provides a low-cost, half-duplex receiver suitable for spectrum monitoring at 2.5 MS/s, whereas the USRP B210 supports programmable transmission with calibrated gain, full-duplex operation, and stable phase coherence required for repeatable interference generation.

\subsubsection{HackRF Configuration}
The HackRF One was controlled through a Python script using the SoapySDR API, which configured center frequency, sampling rate, and analog gain. The receiver continuously sampled the active Wi-Fi channel at a sampling rate of $F_s = 2.5$ MS/s. The resulting in-phase and quadrature (IQ) stream was transformed using a 512-point fast Fourier transform (FFT) to obtain power spectral density (PSD) estimates over a 20 MHz observation window centered on the active Wi-Fi channel. From each PSD frame, five spectral descriptors were computed: mean PSD, standard deviation of PSD, maximum PSD, spectral entropy, and channel occupancy. After each band switch, the HackRF was retuned to align with the active communication channel, and a five-sample cooldown interval was applied before detection resumed to discard stale or buffered measurements.

\subsubsection{USRP B210 Configuration}
The NI USRP B210 was controlled through a Python script using the UHD library, enabling programmable selection of center frequency, modulation type, and output gain. The USRP generated narrowband continuous-wave (CW) tones within the targeted Wi-Fi channel to emulate intentional electromagnetic interference. For 2.4 GHz trials, the jammer transmitted a CW tone at 2.447 GHz with a 20 MHz observation window for the detector; for 5 GHz trials, the jammer transmitted at 5.2 GHz. The USRP transmit gain was fixed at 85 dB for all trials, and the interference duration was fixed at 5s per event.

\subsection{Wireless Network Topology}
The robot operated within a dual-band wireless environment in which a single router exposed two basic service set identifiers (BSSIDs): one operating in the 2.4 GHz band and one in the 5 GHz band. A third 6 GHz network was advertised to the host workstation but was not supported by the Raspberry Pi 4 wireless interface and was therefore excluded from the experiments. This configuration enabled repeated evaluation of robot behavior under interference on either band while keeping the access-point hardware constant across all trials, eliminating router heterogeneity as a confounding variable.

\subsection{Switching Architectures}
Three switching architectures were implemented and compared. The first two use a single Wi-Fi interface with software-based re-association, while the third uses two pre-associated Wi-Fi interfaces with hardware-assisted route switching.

\subsubsection{\emph{Static-Threshold Software Switching (Baseline)}}

The baseline architecture used a single Wi-Fi interface and a HackRF-based static-threshold jamming detector to initiate software-based band re-association. The detector monitored a 20 MHz observation window centered on the active Wi-Fi channel and declared a jamming event when the detector metric exceeded a fixed threshold of -30 dB for 2.4 GHz operation or -18 dB for 5 GHz operation. To suppress false triggering from transient fluctuations, the detector required $N_{sample}=75$ consecutive above-threshold samples before recording an event.

Once a valid jamming event was declared, the robot initiated a full reconnect cycle to the alternate Wi-Fi band using \textit{wpa supplicant}. This process required the wireless interface to deauthenticate from the current access point, scan for the target SSID, authenticate to the new BSSID, complete the WPA2 four-way handshake, restore IP-layer connectivity, and re-establish application-layer ROS 2 communication. Because the same wireless interface had to fully leave one network before joining the other, this architecture incurred the full latency of the re-association process.

\subsubsection{\emph{Adaptive ML-Based Software Switching}}

The second architecture retained the same single-interface software re-association mechanism but replaced the static threshold with a machine learning–based adaptive detector. The five spectral descriptors described in Section IV-B.1 (mean PSD, standard deviation of PSD, maximum PSD, spectral entropy, and channel occupancy) were extracted from each FFT frame and supplied to a Random Forest classifier. The classifier was trained offline on labeled recordings containing both nominal Wi-Fi traffic and active IEMI conditions collected at both operating frequencies. 

At inference time, the classifier produced a probability score representing the likelihood of an active jamming event. This score was compared against a band-specific decision threshold, set to 75\% for 2.4 GHz operation and 50\% for 5 GHz operation. The lower threshold for the 5 GHz band reflected the lower observed signal-to-noise ratio in that operating condition. To suppress false positives arising from transient spectral fluctuations, the detector required three consecutive positive classifications before declaring a jamming event. Following each detection, a five-sample cooldown interval was applied so that the HackRF could retune and clear any buffered samples before normal detection resumed. Once a jamming event was confirmed, the robot executed the same software-driven re-association sequence used in the static-threshold baseline, and therefore incurred the same link-layer handshake latency. Consequently, this architecture isolates the contribution of detector adaptivity while holding the recovery path constant relative to the baseline.

\subsubsection{\emph{Adaptive Hardware-Assisted Switching}} 

The third architecture combined the same adaptive ML-based detector with a hardware-assisted switching mechanism. The detection pipeline operated identically to the configuration described in Section IV-D.2, including the FFT parameters, the five-feature descriptor set, the Random Forest classifier, the band-specific decision thresholds, the three-consecutive-detection requirement, and the post-detection cooldown. 

Rather than relying on a single interface to deauthenticate and re-associate between BSSIDs, the robot maintained two active Wi-Fi interfaces simultaneously. A TP-Link AC600 USB adapter was permanently associated with the 2.4 GHz BSSID, while the onboard wireless interface was permanently associated with the 5 GHz BSSID. Both links remained authenticated and active at all times in a hot-standby configuration, and each interface was assigned a static IP address on the same subnet. When a jamming event was declared, the system redirected outbound telemetry and control traffic by modifying the Linux default route, thereby steering traffic through the unaffected interface without disconnecting from the jammed band. Because both links remained continuously established, this architecture bypassed the full re-association cycle required by the software-only methods and compressed the recovery interval into the millisecond range.

\begin{figure}[!htbp]
  \centering
  \includegraphics[width=0.99\columnwidth]{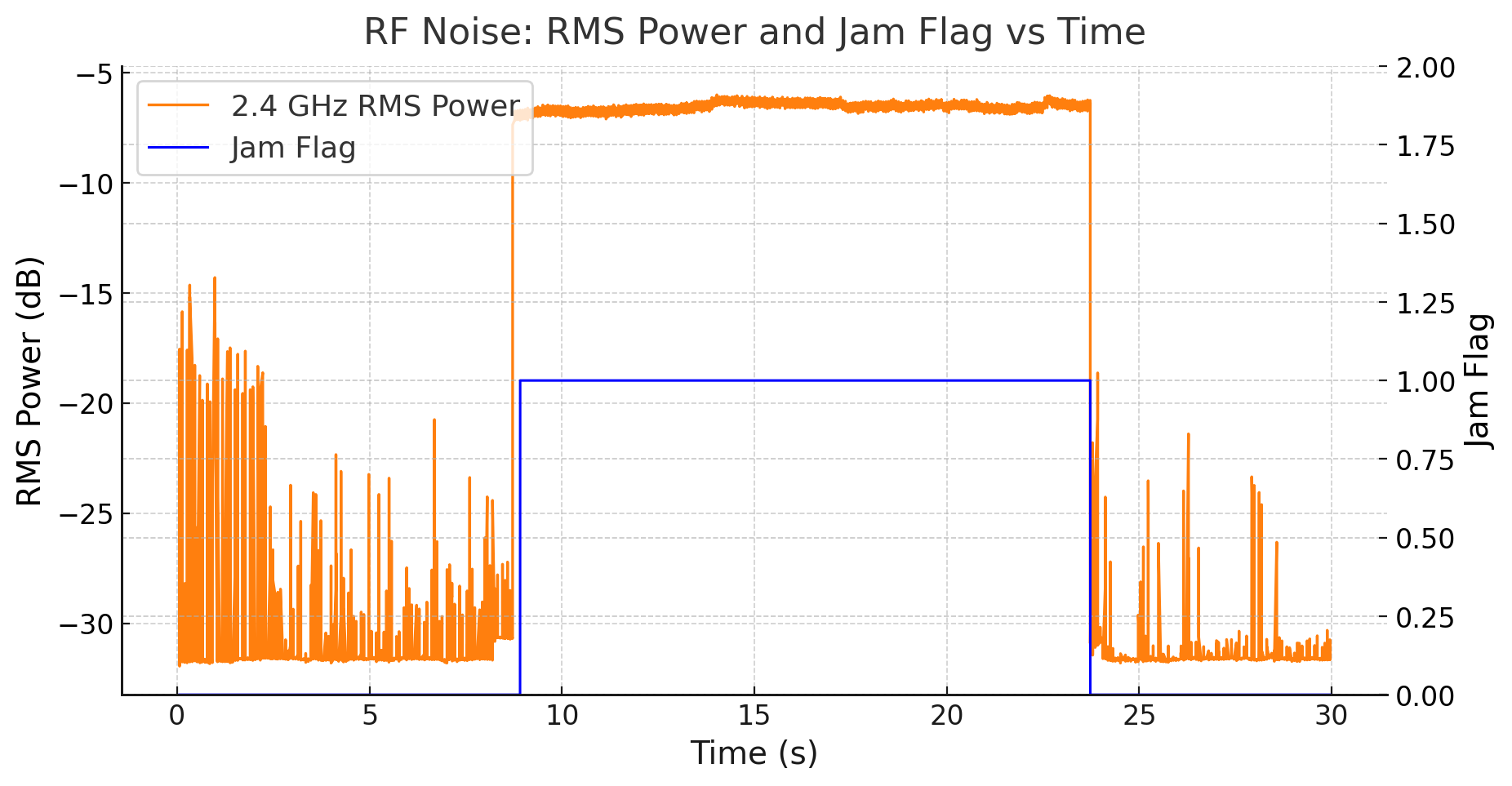}
  \caption{2.4 GHz RMS power and jam-flag output vs. time.}
  \label{fig:jam_flag}
\end{figure}

\section{Experiments and Results}
\begin{figure*}[!t]
  \centering
  % Left Side: Large Spatial Plot (takes full height)
  \begin{subfigure}[b]{0.48\linewidth}
    \centering
    \includegraphics[width=0.7\textwidth]{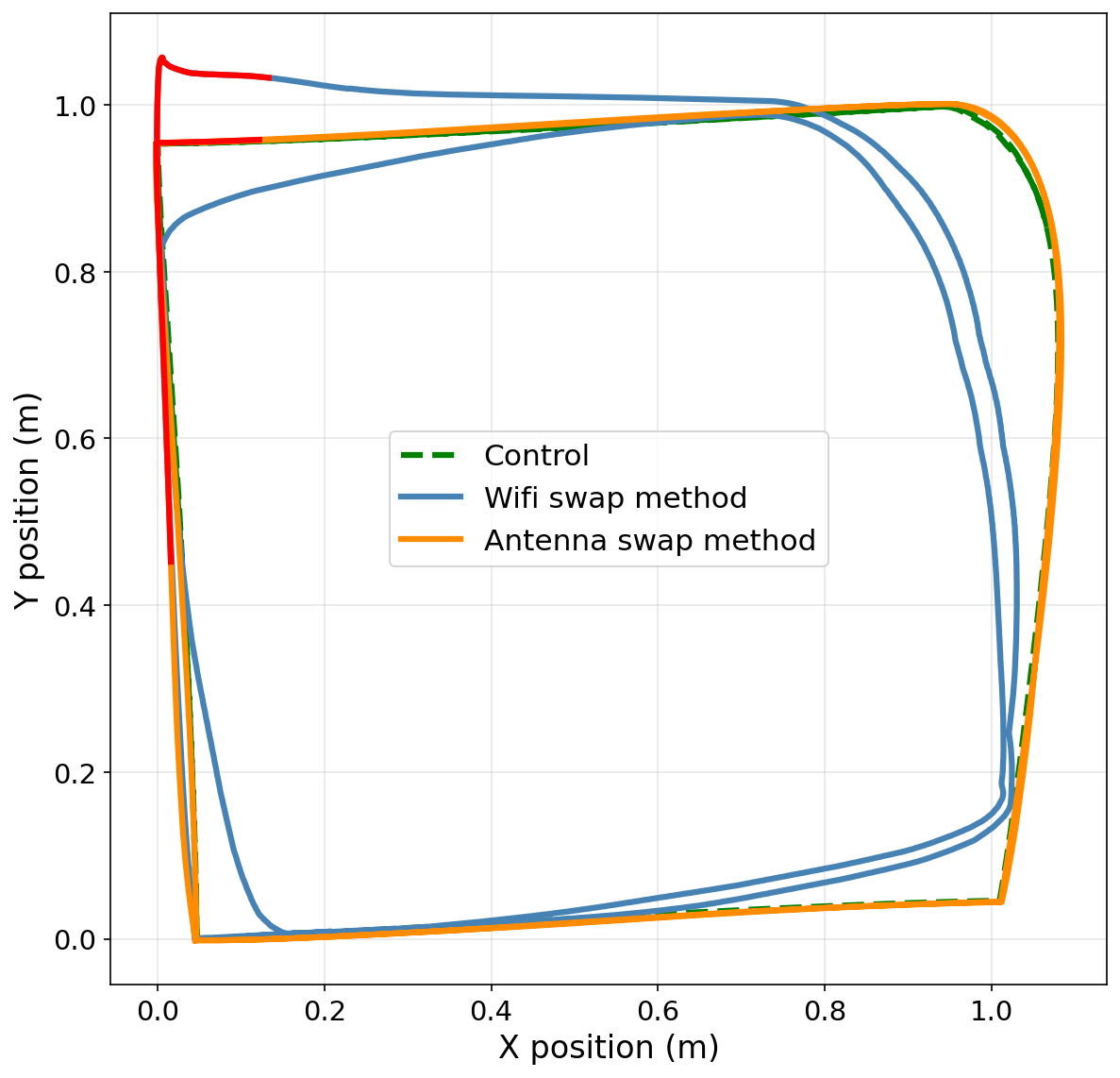}
    \caption{Spatial $X$-$Y$ Trajectory Traces}
    \label{fig:xyplot}
  \end{subfigure}
  \hfill
  % Right Side: Stacked Time-Domain Plots via minipage
  \begin{minipage}[b]{0.48\linewidth}
    \centering
    \begin{subfigure}[b]{\textwidth}
      \centering
      \includegraphics[width=0.7\textwidth]{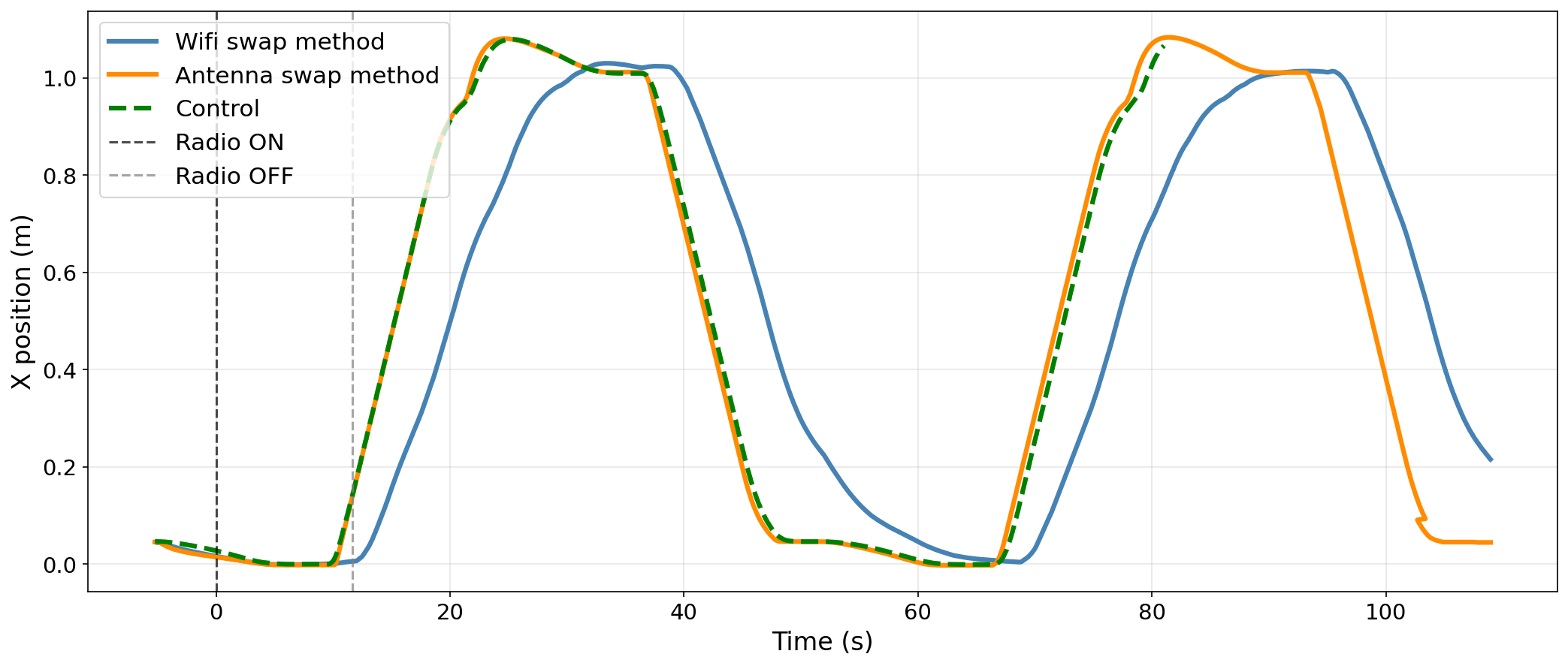}
      \caption{$X$ Position vs. Time}
      \label{fig:xplot}
    \end{subfigure}
    \vspace{0.5cm} % Adjust vertical spacing between the two right stacked plots
    \begin{subfigure}[b]{\textwidth}
      \centering
      \includegraphics[width=0.7\textwidth]{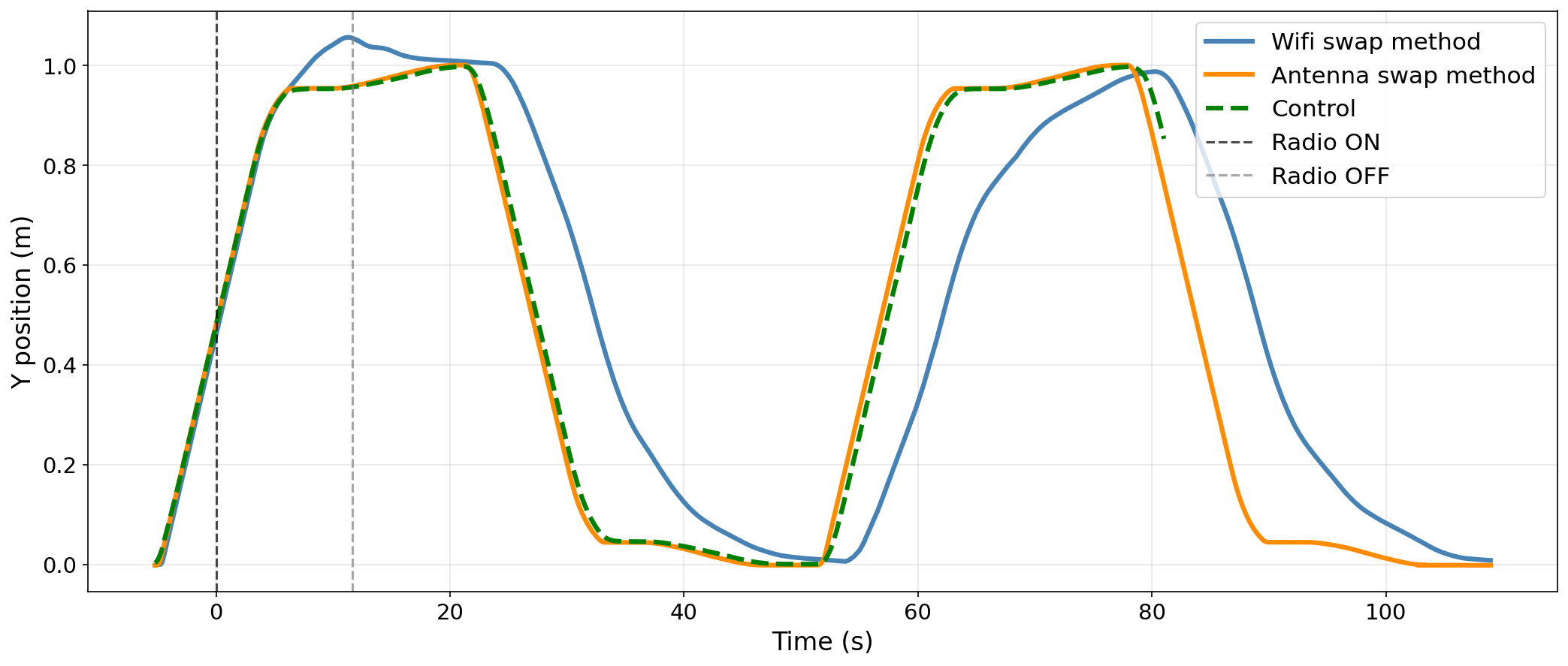}
      \caption{$Y$ Position vs. Time}
      \label{fig:yplot}
    \end{subfigure}
  \end{minipage}
  \caption{Empirical kinematic trajectory results across control, adaptive software switching, and adaptive hardware-assisted switching loops under active IEMI constraints.}
  \label{fig:combined_trajectories}
\end{figure*}

The following experiments were designed to evaluate the impact of intentional RF interference on real-time network recovery and closed-loop waypoint-tracking performance, while directly comparing the three proposed switching architectures. Baseline characterization under nominal conditions revealed distinct spectral profiles for the 2.4~GHz and 5~GHz bands, confirming that static power thresholds are highly susceptible to false positives due to environmental noise. To mitigate this, the trained Random Forest classifier was utilized to evaluate jamming probabilities against band-specific thresholds (75\% for 2.4~GHz; 50\% for 5~GHz), enforcing a strict persistence check of three consecutive positive intervals before declaring an Intentional Electromagnetic Interference (IEMI) event.

All experiments were executed using a stationary NI USRP B210 configured as a continuous-wave (CW) jammer (transmit gain fixed at 85~dB) and a robot-side HackRF front-end detector. The TurtleBot3 Burger executed two continuous laps of a 1~m $\times$ 1~m square path (60~s per lap). This trajectory subjected the platform to a variable robot-to-transmitter separation distance ranging from 2~m to 4~m. The jammer was programmed to activate 8~s into the first leg of the first lap, sustaining targeted in-band interference for 5~s. Telemetry was logged locally to prevent data loss during link outages.

\subsection{Interference Detection Validation}
\begin{table}[h]
\centering
\caption{Interference Target and Sensing Parameters}
\label{table:int_params}
\begin{tabular}{l c}
\toprule
\textbf{Parameter} & \textbf{Value} \\ 
\midrule
Center Frequency & 2.447 GHz / 5.200 GHz \\ 
Interference Waveform & Narrowband CW Tone \\ 
USRP Transmit Gain & 85 dB \\
Emission Duration & 5.0 s \\
SDR Sampling Rate ($F_s$) & 2.5 MS/s \\ 
Sustained Filter Window ($N_{\text{samples}}$) & 75 samples \\  
\bottomrule
\end{tabular}
\end{table}

The initial test validated detection accuracy under a controlled jamming envelope outlined in Table~\ref{table:int_params}. When the CW interference source was engaged, the localized sensing engine observed a rapid escalation in channel energy. Guided by the sample-bound latency model in Eq.~\ref{eq:Delta}, requiring $N_{\text{samples}} = 75$ consecutive above-threshold spectral readings at an acquisition rate of 2.5~MS/s yielded a deterministic detection latency of approximately 0.6~seconds. Empirical measurements perfectly matched this analytical prediction, validating the baseline reliability of the sensing front-end.
\begin{equation}
\Delta = \frac{1}{F_s} \cdot N_{\text{samples}}
\label{eq:Delta}
\end{equation}
The real-time behavior of this detection pipeline is illustrated in Fig.~\ref{fig:jam_flag}, which tracks the measured 2.4~GHz root-mean-square (RMS) power alongside the binary jam-flag output across a representative experimental run. Under nominal conditions, the ambient background noise exhibits low-amplitude fluctuations well below the classification boundary. Upon jammer activation at approximately $t = 8$~seconds, the incoming RMS signal power experiences a sharp step-increase of over 20~dB.

\subsection{Channel Switching Latency Analysis}
Communication recovery performance was assessed by measuring the exact temporal interval between threat declaration and network restoration (defined by a successful end-to-end ICMP ping verification to the access point). 

\begin{figure}[!h]
  \centering
  \includegraphics[width=0.6\columnwidth]{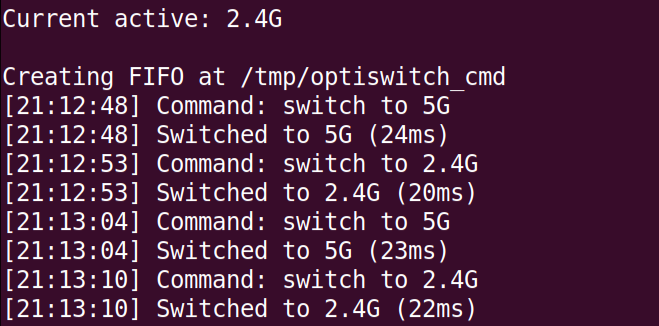}
  \caption{Capture of successful network swaps}
  \label{fig:screenshot}
  \vspace{-2em}
\end{figure}

% \vspace{-2em}
\begin{table}[h]
\caption{Empirical Communication Recovery Latency}
\label{table:channel_switching}
\centering
\begin{tabular}{lc}
\toprule
\textbf{Switching Architecture} & \textbf{Average Recovery Time (s)} \\ 
\midrule
Static-Threshold Software Switching   & 6.001 \\
Adaptive-Threshold Software Switching & 2.549 \\
Adaptive Hardware-Assisted Switching  & 0.141 \\ 
\bottomrule
\end{tabular}
\end{table}

As summarized in Table~\ref{table:channel_switching}, the software-based approaches incurred multi-second connection bottlenecks[cite: 198]. The static-threshold software framework required an average of 6.001~s to recover, heavily penalized by false triggers and single-interface reassociation overhead. Upgrading to machine learning-based feature detection reduced this recovery window to 2.549~s by optimizing band-selection choices and eliminating false-positive switches. However, the proposed Adaptive Hardware-Assisted Switching framework bypassed the link-layer handshake entirely by maintaining dual active, pre-authenticated interfaces. By migrating failover execution to a network-layer routing table modification, it compressed average communication downtime to just 141~ms.

\subsection{Closed-Loop Robotics Path Tracking Impact}
To understand how physical-layer interference propagates to application-layer motion integrity, position telemetry was cross-examined across 15 distinct runs. Path-following accuracy was benchmarked via pooled root-mean-square (RMS) tracking error relative to an unjammed control trajectory.

% \begin{table}[t]
% \caption{Path Tracking Errors and Run Deviations}
% \label{tab:tracking_performance}
% \centering
% \small
% \begin{tabular}{l ccc}
% \toprule
% \textbf{Switching Method} & \textbf{Pooled RMS (cm)} & \textbf{Best Lap (cm)} & \textbf{Worst Lap (cm)} \\ 
% \midrule
% Adaptive Software  & 2.84  & 0.70 (Lap 12)  & 4.37 (Lap 8) \\
% Adaptive Hardware-Assisted  & 0.60  & 0.32 (Lap 2)  & 0.88 (Lap 3)  \\ 
% \bottomrule
% \end{tabular}
% \end{table}

\begin{table}[h]
\caption{Path Tracking Errors and Run Deviations}
\label{tab:tracking_performance}
\centering
\scriptsize
\begin{tabular}{l ccc}
\toprule
\textbf{Switching Method} & \textbf{\makecell{Pooled\\RMS (cm)}} & \textbf{\makecell{Best Lap\\(cm)}} & \textbf{\makecell{Worst Lap\\(cm)}} \\ 
\midrule
Adaptive Software          & 2.84 & \makecell{0.70\\(Lap 12)} & \makecell{4.37\\(Lap 8)} \\ \addlinespace
Adaptive Hardware-Assisted & 0.60 & \makecell{0.32\\(Lap 2)}  & \makecell{0.88\\(Lap 3)}  \\ 
\bottomrule
\end{tabular}
\end{table}

% \vspace{-1em}
The kinematic tracking data in Table~\ref{tab:tracking_performance} reveals a profound divergence in closed-loop stability. Under Adaptive Software Switching, the multi-second loss of velocity commands caused severe path drift, yielding a pooled RMS error of 2.84~cm and an unacceptably high worst-case lap deviation of 4.37~cm. Conversely, the sub-second recovery of the Adaptive Hardware-Assisted routing framework insulated the ROS~2 middleware navigation loop from significant packet starvation. This approach achieved a pooled RMS path error of just 0.60~cm—representing a 78.9\% reduction in trajectory error relative to software switching. Crucially, its worst-case trajectory deviation (0.88~cm) outperformed even the best-case software switching response (0.70~cm), highlighting superior operational consistency.

% \begin{figure*}[!t]
%   \centering
%   \begin{subfigure}[b]{0.32\linewidth}
%     \centering
%     \includegraphics[width=\textwidth]{final_xy_trajectory.png}
%     \caption{Spatial $X$-$Y$ Trajectory Traces}
%     \label{fig:xyplot}
%   \end{subfigure}
%   \hfill
%   \begin{subfigure}[b]{0.32\linewidth}
%     \centering
%     \includegraphics[width=\textwidth]{final_x_vs_time.png}
%     \caption{$X$ Position vs. Time}
%     \label{fig:xplot}
%   \end{subfigure}
%   \hfill
%   \begin{subfigure}[b]{0.32\linewidth}
%     \centering
%     \includegraphics[width=\textwidth]{final_y_vs_time.png}
%     \caption{$Y$ Position vs. Time}
%     \label{fig:yplot}
%   \end{subfigure}
%   \caption{Empirical kinematic trajectory results across control, adaptive software switching, and adaptive hardware-assisted switching loops under active IEMI constraints.}
%   \label{fig:combined_trajectories}
% \end{figure*}

These tracking improvements are visually apparent in the spatial and time-domain kinematic plots grouped in Fig.~\ref{fig:combined_trajectories}. While the control trajectory traces a sharp, nominal square path, the software-switching case experiences severe distortion during and immediately following the 5-second jamming window. In contrast, the hardware-assisted failover response tracks the nominal baseline smoothly, proving that mitigating communication downtime at the network layer prevents physical-layer adversarial attacks from cascading into system-level task failure.

\section{Conclusion and Future Work}\label{sec:futurework}
 
This work presented an empirical evaluation of intentional electromagnetic
interference (IEMI) on a ROS 2-based autonomous mobile robot, demonstrating
that communication-failover latency dictates physical task integrity. By
coupling a software-defined radio sensing front-end with a Random Forest
classifier, the proposed Adaptive Hardware-Assisted Switching framework
bypassed link-layer handshake bottlenecks by executing failover at the network
layer, considerably improving average recovery latency (141~ms) and reducing
pooled root-mean-square (RMS) path error (78.9\%) relative to software
re-association. Ultimately, these findings verify that mitigating communication
disruptions at the network layer prevents physical-layer adversarial attacks
from cascading into system-level mission failure. The present evaluation is limited to a continuous-wave jammer on a single robotic platform, leaving more capable adversaries untested. Future work will accordingly broaden the threat model to reactive, protocol-aware, and frequency-agile jammers that can follow the link across bands, and extend the detector and failover policy to remain effective against them.
 
\color{black}

\section*{Acknowledgment}
This work was supported in part by Computing Research Association (CRA) UR2PHD program. Gemini Generative AI tools were employed in refining the technical narrative throughout the manuscript. 

\bibliographystyle{IEEEtran}
\bibliography{references}
\end{document}